# Efficient MIMO-OFDM Schemes for Future Terrestrial Digital TV with Unequal Received Powers

Youssef Nasser *member IEEE*, Jean-François Hélard *Senior member IEEE*, Mathieu Crussière and

Oudomsack Pasquero

*Institute of Electronics and Telecommunications of Rennes, UMR CNRS 6164, Rennes, France*

*Email :* youssef.nasser@insa-rennes.fr

*Abstract*- **This article investigates the effect of equal and unequal received powers on the performances of different MIMO-OFDM schemes for terrestrial digital TV. More precisely, we focus on three types of non-orthogonal schemes: the BLAST scheme, the Linear Dispersion (LD) code and the Golden code, and we compare their performances to that of Alamouti scheme. Using two receiving antennas, we show that for moderate attenuation on the second antenna and high spectral efficiency, Golden code outperforms other schemes. However, Alamouti scheme presents the best performance for low spectral efficiency and equal received powers or when one antenna is dramatically damaged. When three antennas are used, we show that Golden code offers the highest robustness to power unbalance at the receiving side.**

*Keywords*- **OFDM, MIMO, Space Time codes.**

## I. INTRODUCTION

The potential advantages of digital television broadcasting over conventional analogue broadcasting are numerous and well known. For broadcasters, digital technology offers significantly improved operational flexibility, providing the means for new services which go beyond the scope of conventional television programmes. Since its inauguration in 1993, digital video broadcast (DVB) project for terrestrial (DVB-T) transmission has fully responded to the objectives of its designers, delivering wireless digital TV services in almost every continent [1]. In fact, there is no single DVB standard, but rather a collection of standards, technical recommendations and guidelines. In Spring 2006, DVB community was asked to provide technical specifications and studies for a future second generation of DVB-T called DVB-T2. It is expected that the first profile of DVB-T2 specification, for fixed reception of high definition television (HDTV) services, will be completed as soon as possible, with a second profile offering improved mobile performance completed around the end of 2008. Against this background, a new European CELTIC project called *Broadcast for 21$^{st}$ Century* (B21C) was launched [2]. It constitutes a contribution task force to the reflections engaged by the DVB project and should give a real support for the conclusions and decisions within DVB project, particularly on multiple input multiple output (MIMO) with orthogonal frequency division multiplexing (OFDM) transmission for HDTV services.

The work presented in this paper has been carried out within the framework of B21C project. The contribution of this work is twofold. First, a generalized framework is proposed for modelling the effect of unequal received powers on different receiving antennas. Therefore, we analyze and compare some of the most promising MIMO-OFDM systems in the context of broadcasting for future terrestrial digital TV with equal but also unequal received powers i.e. with unequal received signal to noise ratio (SNR) per antenna. In the literature, most of the works consider equal received powers for the performance comparison of MIMO-OFDM schemes [3][4]. The assumption of unequal received powers could be seen in different communications contexts like in a broadcast link where two different antennas are used at the receiving side or in a mobile link. Indeed, the call for technology within DVB-T2 consortium moves towards an expectation of such situations where one outdoor antenna (roof antenna for example) and one or two indoor antennas are used. Eventually, we note that for complexity reasons the analysis of different MIMO-OFDM systems is not achieved with the optimal maximum likelihood (ML) receiver. Instead, we use a sub-optimal iterative receiver with few iterations.

This paper is structured as follows. Section 2 describes the system model for MIMO-OFDM. In section 3 we discuss the choice of different MIMO schemes considered in this paper. Section 4 presents the iterative receiver with a detailed description of its blocks. Simulation results are drawn in section 5. Section 6 concludes the paper.

## II. SYSTEM MODEL WITH UNEQUAL RECEIVED POWERS

Consider an OFDM communication system using $M_T$ transmit antennas (Tx) and $M_R$ receive antennas (Rx) for a downlink communication. Such a system could be implemented for the $M_T$ transmit antennas using a space-time (ST) encoder which takes $Q$ data complex symbols and transforms them to a ($M_T$,$T$) output matrix according to the ST block coding (STBC) scheme. The ST STBC coding rate is then defined by $L=Q/T$. Figure 1 depicts the transmitter modules. Information bits $b_k$ are first channel encoded with a convolutional encoder of coding rate $R$, randomly interleaved, and fed directly to a quadrature amplitude modulation (QAM) module which assigns $B$ bits for each of the complex constellation points. Therefore, each group $\mathbf{s}=[s_1,\ldots,s_Q]$ of $Q$ complex symbols $s_q$ becomes the input of the STBC encoder. Let $\mathbf{X}=[x_{i,t}]$ where $x_{i,t}$ ($i=1,\ldots,M_T$; $t=1,\ldots,T$) be the output of STBC encoder. This output is then fed to $M_T$ OFDM modulators, each using N subcarriers.

In order to have a fair analysis and comparison between different STBC codes, the signal power at the output of the ST encoder is normalized by $M_T$. We assume in this work that the transmission from a transmitting antenna $i$ and a receiving antenna $j$ is achieved for each subcarrier $n$ through a frequency non-selective Rayleigh fading channel. The latter is assumed to be constant during $T$ symbol durations. With these assumptions, the channel coefficients $h_{i,j}[n]$ are assumed as independent complex Gaussian distributed samples with unit variance. We assume also that the transmitter and receiver are perfectly synchronised. Moreover, we assume perfect channel state information (CSI) at the receiver.

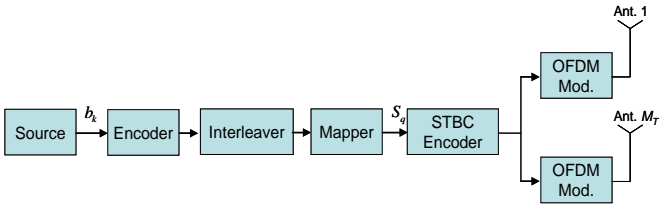

Figure 1- MIMO-OFDM transmitter.

Since we assume a frequency domain transmission, the signal received on the subcarrier $n$ by the antenna $j$ is a superposition of the transmitted signal by the different antennas multiplied by the channel coefficients to which white Gaussian noise (WGN) is added. It is given by:

$$y_j[n,t] = \sqrt{\alpha_j} \sum_{m=1}^{M_T} h_{i,j}[n] x_i[n,t] + w_j[n,t] \quad (1)$$

where $y_j[n,t]$ is the signal received on the $n^{th}$ subcarrier by the $j^{th}$ receiving antenna during the $t^{th}$ OFDM symbol duration. $h_{i,j}[n]$ is the frequency channel coefficient assumed to be constant during $T$ symbol durations, $x_i[n,t]$ is the signal transmitted by the $i^{th}$ antenna and $w_j[n,t]$ is the additive WGN with zero mean and variance $N_0/2$. $\alpha_j$ is the power attenuation factor of the $j^{th}$ receiving antenna. By introducing an equivalent receive matrix $\mathbf{Y}[n] \in C^{M_R \times T}$ whose elements are the complex received symbols expressed in (1), we can write the received signal on the $n^{th}$ subcarrier on all receiving antennas as:

$$\mathbf{y}[n] = \mathbf{AH}[n]\mathbf{X}[n] + \mathbf{W}[n] \quad (2)$$

Where $\mathbf{H}[n]$ is the $(M_R, M_T)$ channel matrix whose components are the coefficients $h_{i,j}[n]$, $\mathbf{X}[n]$ is a $(M_T, T)$ complex matrix containing transmitted symbols $x_i[n,t]$. $\mathbf{W}[n]$ is a $(M_R, T)$ complex matrix corresponding to the WGN. Since we assume unequal received powers, $A$ is a $(M_R, M_R)$ diagonal matrix whose diagonal elements are the square roots of the power attenuation factors $\alpha_j$ associated to each receiving antenna. Without loss of generality, we will drop the subcarrier index $n$ in the sequel.

Let us now describe the transmission link with a general model independently of the ST coding scheme. We separate the real and imaginary parts of the entries $s_q$, of the outputs $\mathbf{X}$ of the ST encoder as well as those of the channel matrix $H$ and the received signal $y$. Let $s_{q,R}$ and $s_{q,I}$ be the real and imaginary parts of $s_q$. The main parameters of the code are given by its dispersion matrices $\mathbf{U_q}$ and $\mathbf{V_q}$ ($q=1,\ldots,Q$), corresponding (not equal) respectively to the real and imaginary parts of $\mathbf{X}$. With these notations, $\mathbf{X}$ is given by:

$$\mathbf{X} = \sum_{q=1}^{Q} \left( s_{q,\Re} \mathbf{U_q} + j s_{q,\Im} \mathbf{V_q} \right) \quad (3)$$

We separate the real and imaginary parts of $\mathbf{S}$, $\mathbf{Y}$ and $\mathbf{X}$ and stack them row-wise in vectors of dimensions $(2Q,1)$, $(2M_RT,1)$ and $(2M_TT,1)$ respectively. We obtain:

$$\mathbf{s} = \left[ s_{1,\Re}, s_{1,\Im}, \ldots, s_{Q,\Re}, s_{Q,\Im} \right]^{tr}$$

$$\mathbf{y} = \left[ y_{1,\Re}, y_{1,\Im}, \ldots, y_{T,\Re}, y_{T,\Im}, \ldots, y_{M_RT,\Re}, y_{M_RT,\Im} \right]^{tr} \quad (4)$$

$$\mathbf{x} = \left[ x_{(1,1),\Re}, x_{(1,1),\Im}, \ldots, x_{(2M_T,T),\Re}, x_{(2M_T,T),\Im} \right]^{tr}$$

where $tr$ holds for matrix transpose.

Since, we use linear ST coding, vector $\mathbf{x}$ can be written as:

$$\mathbf{x} = \mathbf{F}.\mathbf{s} \quad (5)$$

where $\mathbf{F}$ has the dimensions $(2M_TT, 2Q)$ and is obtained through the dispersion matrices of the real and imaginary parts of $\mathbf{s}$. It is given by:

$$\mathbf{F} = \begin{bmatrix} \mathbf{F_1}(1,1) & \cdots & \cdots & \mathbf{F_Q}(1,1) \\ \vdots & \vdots & \vdots & \vdots \\ \mathbf{F_1}(1,T) & \cdots & \cdots & \mathbf{F_Q}(1,T) \\ \vdots & \ddots & \vdots & \vdots \\ \mathbf{F_1}(M_T,T) & \cdots & \cdots & \mathbf{F_Q}(M_T,T) \end{bmatrix} \quad (6)$$

$\mathbf{F}$ is composed of $M_T$ blocks of $2T$ rows each i.e. the data transmitted on each antenna is gathered in one block having $2T$ rows and $2Q$ columns according to the ST coding scheme. The different components of $\mathbf{F}$ are given by:

$$\mathbf{F_q}(m,t) = \begin{bmatrix} \mathbf{U_{q,R}}(m,t) & -\mathbf{V_{q,I}}(m,t) \\ \mathbf{U_{q,I}}(m,t) & \mathbf{V_{q,R}}(m,t) \end{bmatrix} \quad (7)$$

As we change the formulation of $\mathbf{S}$, $\mathbf{Y}$ and $\mathbf{X}$ in (4), it can be shown that vectors $\mathbf{X}$ and $\mathbf{Y}$ are related through the matrix $\mathbf{G}$ of dimensions $(2M_RT, 2M_TT)$ such that:

$$\mathbf{Y} = \mathbf{BGX} + \mathbf{W} \quad (8)$$

where $\mathbf{B}$ is a $(2M_RT, 2M_RT)$ diagonal matrix whose elements are related to the power attenuations factors by:

$$B_{i,i} = \sqrt{\alpha_j} \quad \begin{array}{l} 2.T(j-1)+1 \leq i \leq 2T.j \\ j = 1,\ldots,M_R \end{array} \quad (9)$$

Matrix **G** is composed of blocks $\mathbf{G}_{i,j}$ ($i=1,…,M_R$; $j=1,…,M_T$) each having ($2T,2T$) elements given by:

$$\mathbf{G}_{i,j} = \begin{pmatrix} h_{(i,j),\Re} & -h_{(i,j),\Im} & 0 & & \cdots & & 0 \\ h_{(i,j),\Im} & h_{(i,j),\Re} & 0 & & \cdots & & 0 \\ 0 & 0 & h_{(i,j),\Re} & -h_{(i,j),\Im} & 0 & \cdots & 0 \\ 0 & 0 & h_{(i,j),\Im} & h_{(i,j),\Re} & 0 & \cdots & 0 \\ 0 & \cdots & 0 & \ddots & 0 & & 0 \\ 0 & \cdots & 0 & & \ddots & 0 & 0 \\ 0 & \cdots & & & 0 & h_{(i,j),\Re} & -h_{(i,j),\Im} \\ 0 & \cdots & & & 0 & h_{(i,j),\Im} & h_{(i,j),\Re} \end{pmatrix}_{(2T,2T)} \quad (10)$$

Now, substituting **x** from (5) in (8), the relation between **y** and **s** becomes:

$$\mathbf{y} = \mathbf{BGFs} + \mathbf{W} = \mathbf{G}_{eq}\mathbf{s} + \mathbf{W} \quad (11)$$

$\mathbf{G}_{eq}$ is the equivalent channel matrix between **s** and **y**. It is assumed to be known perfectly at the receiving side.

### III. CHOICE OF ST SCHEMES

*A. Relation between probability of error and channel capacity*

Assume the channel is totally unknown at the transmitter and perfectly known at the receiver, the optimum power distribution strategy is to allocate equal power over all the subchannels in different domains (time, frequency and space).

Based on (11) and keeping in mind that the channel coefficients of the matrix $\mathbf{G}_{eq}$ in (11) are separated into real and imaginary parts and they are assumed to be constant during $T$ OFDM symbols, the channel capacity of such transmission and a given transmitted power is:

$$C_G = \frac{1}{2T}\log_2 \det\left(\mathbf{I}_{2M_RT} + \mathbf{G}_{eq}\mathbf{R}_{SS}\mathbf{G}_{eq}^{\mathbf{H}}\mathbf{R}_{WW}^{-1}\right) \quad (12)$$

where $\mathbf{R}_{SS}$, and $\mathbf{R}_{WW}$ are the autocorrelation matrix of the data entries **s** and the WGN respectively. We show that the channel capacity is given by:

$$C_G = \frac{1}{2T}\log_2 \det\left(\mathbf{I}_{2M_RT} + \frac{P_0}{\sigma_W^2}\mathbf{BGFF}^{\mathbf{H}}\mathbf{G}^{\mathbf{H}}\mathbf{B}^{\mathbf{H}}\right) \quad (13)$$

And the mean channel capacity over the channel realizations is:

$$C = E[C_G] \quad (14)$$

For non-orthogonal (NO) schemes, the choice of an optimal ST coding matrix depends on some criteria. It is based on an optimization of the pair wise error probability (PEP) or channel capacity and diversity, or a compromise between them. Based on the knowledge of the possible set of matrix **X**, Tarokh [5] proposed some criteria to construct ST coding matrix **X**. In [6], Hassibi is based on the PEP for Gaussian distributed inputs to define a new ST code. The PEP criterion, based initially on ML detection, should be studied further. It consists in minimizing the quantity:

$$\Pr(X \rightarrow X') \leq \frac{1}{2} E\left[\det\left(\mathbf{I}_{2M_RT} + \gamma_x \mathbf{G}_{eq}\mathbf{G}_{eq}^{\mathbf{tr}}\right)^{-1/2}\right] \quad (15)$$

where $\gamma_x$ is the signal to noise ratio (SNR) for each transmitted symbol x∈**X**.

The transfer to the mean error probability is difficult from (15) since there is a large number of matrices **X** which verify the PEP minimization. However, a good issue consists in maximizing the determinant. Surprisingly, the maximization of the determinant in (15) is equivalent to the maximization of channel capacity in (14)[1]. That is, [6] proposes LD scheme based on maximization of channel capacity. This allows imposing some constraints on the choice of dispersion matrix **F**. Since the channel is unknown at the transmit side, the first constraint is to have trace ($\mathbf{F}^{tr}.\mathbf{F}$)=$2T$. The second constraint is to have a uniform repartition of signal power on different transmit antennas. This could be achieved by fulfilling $F$ conveniently. Another interesting point in this analysis consists in the relation between probability of error and capacity. Indeed, 2 different ST schemes have the same channel capacity. However, they present different probabilities of errors since the PEP is upper bounded by (not equal to) a function of the channel capacity inverse. It is shown in [6] that two dispersion matrices having the same channel capacity do not have the same error rate. This is due to the fact that the diversity introduced by the dispersion matrices is different from a code to another.

*B. Considered ST Coding schemes*

As a consequence of the discussion in previous section, we consider in this paper some of the most promising MIMO schemes having the same rate. Therefore for equal spectral efficiencies, (14) shows that all these schemes have the same channel capacity. We will show by simulations in next sections that even with equal channel capacities and SNRs, the probability of error of different schemes is not the same since they have not the same diversity order.

First, we consider the simplest orthogonal ST coding scheme proposed by Alamouti [7] as a reference of comparison. Since $M_T=2$, we have $Q=T=2$ and the ST coding rate $L=1$. This code is given by the matrix:

$$\mathbf{X} = \begin{bmatrix} s_1 & s_2 \\ -s_2^* & s_1^* \end{bmatrix} \quad (16)$$

For non-orthogonal schemes, we consider in this work the well-known multiplexing scheme i.e. the V-BLAST [8]. VBLAST is designed to maximize the rate by transmitting symbols sequentially on different antennas. Its coding scheme is given for $T=1$, $Q=2$ and $L=2$ by:

$$\mathbf{X} = \begin{bmatrix} s_1 & s_2 \end{bmatrix}^{tr} \quad (17)$$

We also consider the LD code proposed by Hassibi [6] for which we have $Q=4$, $T=2$ and $L=2$. It is designed to maximize the mutual information between transmitter and receiver. It is defined by:

---
[1] The inverse does not imply a minimization of PEP.

$$\mathbf{X} = \frac{1}{\sqrt{2}} \begin{bmatrix} s_1 + s_3 & s_2 - s_4 \\ s_2 + s_4 & s_1 - s_3 \end{bmatrix} \quad (18)$$

Finally, we consider the optimized Golden code [9] denoted hereafter by GC. The Golden code is designed to maximize the rate such that the diversity gain is preserved for an increased signal constellation size. It is defined for $Q=4$, $T=2$ and $L=2$ by:

$$\mathbf{X} = \frac{1}{\sqrt{5}} \begin{bmatrix} \beta(s_1 + \theta s_2) & \beta(s_3 + \theta s_4) \\ \mu\overline{\beta}(s_3 + \overline{\theta} s_4) & \overline{\beta}(s_1 + \overline{\theta} s_2) \end{bmatrix} \quad (19)$$

where $\theta = \frac{1+\sqrt{5}}{2}$, $\overline{\theta} = 1-\theta$, $\beta = 1 + j(1-\theta)$, $\overline{\beta} = 1 + j(1-\overline{\theta})$, $\mu = j$ and $j = \sqrt{-1}$.

## IV. ITERATIVE SPACE-TIME RECEIVER

In the case of OSTBC, optimal receiver is made of a concatenation of ST decoder and channel decoder modules (preceded by a bit deinterleaver). In NO-STBC schemes, there is an inter element interference (IEI) at the receiving side. The optimal receiver in this case is based on joint ST and channel decoding operations. However such receiver is extremely complex to implement and requires large memory to store the different points of the trellis. Moreover, it could not be implemented reasonably in one ship. Thus the sub-optimal solution proposed here consists of an iterative receiver where the ST detector and channel decoder exchange extrinsic information in an iterative way until the algorithm converges. The iterative detection and decoding exploits the error correction capabilities of the channel code to provide improved performance. This is achieved by iteratively passing soft *a priori* information between the detector and the soft-input soft-output (SISO) decoder [10].

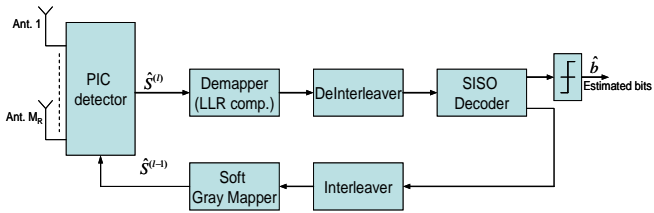

Figure 2- Iterative receiver structure.

### A. STBC detection

In the literature, different detection strategies are presented. The detection problem is to find the transmitted data **s** given the vector **y**. The iterative detector shown in Figure 2 is composed of a MIMO equalizer, a demapper which is made up of a parallel interference cancellation (PIC), a log likelihood ratio (LLR) computation, a soft-input soft-output (SISO) decoder, and a soft mapper.

At the first iteration, the demapper takes the estimated symbols $\hat{\mathbf{s}}$, the knowledge of the channel $\mathbf{G_{eq}}$ and the noise variance, and computes the LLR values (soft information) of each of the $B$ coded bits transmitted per channel use. The estimated symbols $\hat{\mathbf{s}}$ are obtained via minimum mean square error (MMSE) filtering according to:

$$\hat{s}_p = \mathbf{g_p^{tr}} \left( \mathbf{G_{eq}} \cdot \mathbf{G_{eq}^{tr}} + \sigma_w^2 \mathbf{I} \right)^{-1} \mathbf{y} \quad (20)$$

where $\mathbf{g_p^{tr}}$ of dimension $(2M_RT, 1)$ is the $p^{th}$ column of $\mathbf{G_{eq}}$ ($1 \leq p \leq 2Q$). $\hat{s}_p$ is the estimation of the real part ($p$ odd) or imaginary part ($p$ even) of $s_q$ ($1 \leq q \leq Q$).

### B. LLR computation

As we consider Gray mapping with QAM modulation of $B$ bits per symbol, the computation of LLR is done as in [11]. We note that we use the approximation of $\log(\exp(x_1)+(\exp(x_2)) \approx \max(x_1,x_2)$. This simplifies considerably the LLR expressions especially for high constellations. We note also that the total noise variance corresponding to the additive WGN and the IEI is used for LLR computation.

### C. SISO decoder

The deinterleaved soft information ($LLR_{k,p}$) of the $k^{th}$ of the $p^{th}$ symbol at the output of the demapper becomes the input of the outer decoder. The outer decoder computes the a *posteriori* information of the information bits and of the coded bits. The *a posteriori* information of the coded bits produces new (and hence) extrinsic information $LLR_{k,p}^{ext}$ of the coded bits upon removal of the *a priori* information[2] and minimizing the correlation between input values $LLR_{k,p}$. In our work, SISO decoding is based on the Max-Log-MAP algorithm [10]. The extrinsic information at the output of the channel decoder is then interleaved and fed to a soft Gray mapper module.

### D. Soft mapper

The soft mapper achieves reciprocal operation of soft demapper. Knowing the extrinsic information of the $k^{th}$ bit of the $q^{th}$ symbol, the soft estimation of the complex symbol $s_q$, noted hereafter $\widetilde{s}_q$, is defined by:

$$\widetilde{s}_q = \mathrm{E}\left\{ s \middle| LLR_{1,q}^{ext}, \ldots, LLR_{B,q}^{ext} \right\} \quad (21)$$

where E holds for expectation function. Let $[c_1,\ldots,c_B]$ the set of bits constituting the constellation point $s$. Equation (21) yields:

$$\widetilde{s}_q = \sum_{s \in \psi} s \cdot \Pr(\widetilde{s}_q = s) \quad (22)$$

where $\psi$ is the set of constellation points and:

$$\Pr(\widetilde{s}_q = s) = \prod_{s:=[c_1,\ldots,c_B]} \Pr(\widetilde{c}_{k,q} = c_k) \quad (23)$$

The probability expressions in (23) are deduced from the LLR expressions as:

---

[2] when the transmitted bits are likely equal, this information is equal to zero.

$$\Pr(\widetilde{c}_{k,q} = 1) = \frac{\exp(LLR_{k,q}^{ext})}{1 + \exp(LLR_{k,q}^{ext})} \quad (24)$$

$$\Pr(\widetilde{c}_{k,q} = 0) = 1 - \Pr(\widetilde{c}_{k,q} = 1)$$

Once the estimation of the different symbols $s_q$ is achieved by the soft mapper at the first iteration, we use this estimation for the next iterations process. From the second iteration, we perform PIC operation followed by a simple inverse filtering (instead of MMSE filtering at the first iteration):

$$\hat{\mathbf{y}}_\mathbf{p} = \mathbf{y} - \mathbf{G}_{\mathbf{eq},\mathbf{p}} \tilde{\mathbf{s}}_\mathbf{p}$$
$$\hat{s}_p = \frac{1}{\mathbf{g}_\mathbf{p}^{\mathbf{tr}} \mathbf{g}_\mathbf{p}} \mathbf{g}_\mathbf{p}^{\mathbf{tr}} \hat{\mathbf{y}}_\mathbf{p} \quad (25)$$

where $\mathbf{G}_{\mathbf{eq},\mathbf{p}}$ of dimension ($2M_R T$, $2Q$-1) is the matrix $\mathbf{G}_{\mathbf{eq}}$ with its $p^{th}$ column removed, $\tilde{\mathbf{s}}_\mathbf{p}$ of dimension ($2Q$-1, 1) is the vector $\tilde{\mathbf{s}}$ estimated by the soft mapper with its $p^{th}$ entry removed.

## V. SIMULATION RESULTS

In this section, we present a comparative study of the different ST coding schemes described in section 3. The performance comparison is made in terms of bit error rate (BER) for the cases of equal and unequal received powers at the receiving side.

We assume that 2 or 3 receiving antennas are used. For equal received powers, we assume that the power attenuation factors of matrix A in (2) are equal to 0dB i.e. $\alpha_1 = \alpha_2 = \alpha_3$ 0dB. For unequal received powers, we set $\alpha_1$ to 0dB and we change $\alpha_2$ and $\alpha_3$ such that $\alpha_2 = \alpha_3$.

The simulations parameters considered in this work are derived from those of DVB-T. They are given in Table 1. The spectral efficiencies $\eta$=2, 4 and 6 [bit/sec/Hz] are obtained for different ST schemes as shown in Table 2.

**Table 1- Simulations Parameters**

| | |
|---|---|
| Number of subcarriers | 2K mode (1705 active subcarriers) |
| Number of Tx antennas | 2 |
| Number of Rx antennas | 2 or 3 |
| Rate R of convolutional code | 1/2, 2/3, 3/4 |
| Polynomial code generator | $(133,171)_o$ |
| Channel estimation | perfect |
| Constellation | QPSK, 16-QAM, 64-QAM, 256-QAM |
| Spectral Efficiencies | $\eta$= 2, 4 and 6 [bit/sec/Hz] |
| Power attenuations factors | $\alpha_1$= 0dB<br>$\alpha_2 = \alpha_3$= -12, -9, -6, -3 and 0dB |

First, let us characterize the behavior of the iterative receiver. Figure 3 provides performance of the Golden code with iterative receiver. The performance is given in terms of BER versus the Eb/N0 ratio for different numbers of iterations. We observe on this figure that the iterative process converges for an Eb/N0 greater than a limit value, which is equal to 6 dB in this case. Moreover, we observe that the convergence of the iterative receiver is reached after 3 iterations which means an acceptable complexity as compared to ML detection. This can be observed with Golden code, but also with LD code and VBLAST scheme. That is, for NO-STBC schemes, we will present in the sequel the performances after 3 iterations only.

For equal received powers and 2Rx ($\alpha_1 = \alpha_2$= 0dB), Figure 4 and Figure 5 compare the different ST coding schemes for $\eta$=4 and $\eta$=6 respectively. These figures show that Golden code presents the best performance with respect to other schemes since it benefits from its full diversity. For equal received powers ($\alpha_1 = \alpha_2 = \alpha_3$=0dB), 3Rx and a spectral efficiency $\eta$=6, NO-STBC schemes outperform Alamouti code as depicted in Figure 6. More precisely, for a BER=$10^{-4}$, the Eb/N0 gain for Golden code is roughly equal to 6 dB compared to Alamouti code.

**Table 2- Different MIMO schemes and efficiencies**

| Spectral Efficiency | ST scheme | ST rate L | Constellation | R |
|---|---|---|---|---|
| $\eta$=2 [bit/Sec/Hz] | Alamouti | 1 | 16-QAM | 1/2 |
| | VBLAST | 2 | QPSK | 1/2 |
| | LD | 2 | QPSK | 1/2 |
| | Golden | 2 | QPSK | 1/2 |
| $\eta$=4 [bit/Sec/Hz] | Alamouti | 1 | 64-QAM | 2/3 |
| | VBLAST | 2 | 16-QAM | 1/2 |
| | LD | 2 | 16-QAM | 1/2 |
| | Golden | 2 | 16-QAM | 1/2 |
| $\eta$=6 [bit/Sec/Hz] | Alamouti | 1 | 256-QAM | 3/4 |
| | VBLAST | 2 | 64-QAM | 1/2 |
| | LD | 2 | 64-QAM | 1/2 |
| | Golden | 2 | 64-QAM | 1/2 |

For unequal received powers, conclusions are different. Using 2 receivers, Figure 7, Figure 8 and Figure 9 depict the Eb/N0 ratio required to obtain a BER=$10^{-4}$ for spectral efficiencies $\eta$=2, 4 and 6 [bit/sec/Hz] respectively and different values of the power attenuation factor $\alpha_2$. For $\eta$=2, Figure 7 shows that Alamouti scheme outperforms the other ST coding schemes. Indeed, the required Eb/N0 to obtain a BER=$10^{-4}$ for Alamouti scheme is less than for other schemes. This superiority increases when the received power on the second antenna decreases and can be explained as follows. For equal received powers ($\alpha_1 = \alpha_2$=0dB), all the ST coding schemes present the same performance. When the received power on the second antenna decreases, the different schemes (with 2 receiving antennas) tend to be ST schemes with only one antenna. However, due to the redundancy included by Alamouti scheme, the loss introduced by the power decrease could be simply recovered by the first antenna at the detriment of half power loss in terms of Eb/N0. That is why the maximal loss of Alamouti scheme is upper-bounded. It is of 3dB when $\alpha_2$ passes from 0dB to -12dB. For NO schemes, the redundancy is les pronounced in the ST matrices, which implies a greater Eb/N0 loss when the power of the second receiving antenna passes from 0dB to -12dB. For higher spectral efficiency i.e. $\eta$=4 or 6 [bit/Sec/Hz], Figure 8 and Figure 9 show that Golden code presents the best performance as long as the power attenuation factor on the second antenna $\alpha_2$ is greater than a limit value. Otherwise, Alamouti scheme presents the best performance. This limit value is of -6dB for $\eta$=4 and -9dB for $\eta$=6. This behavior can be explained by the fact that Golden code is designed to maximize the diversity for high signal constellations and equal received powers. The diversity gain is however lost when one antenna is quite turned off.

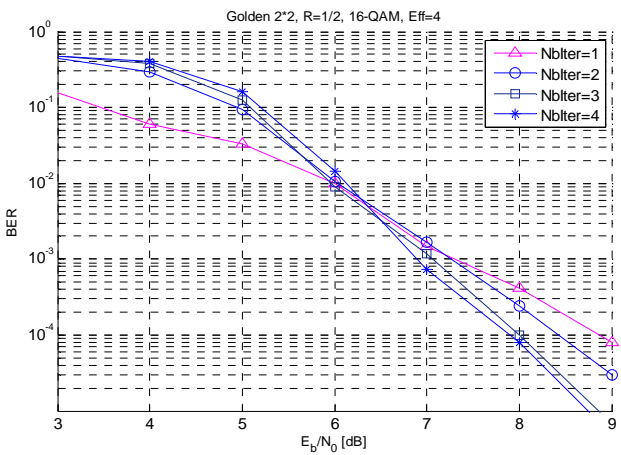

Figure 3- Convergence of Golden code with respect to the number of iterations, 2Rx.

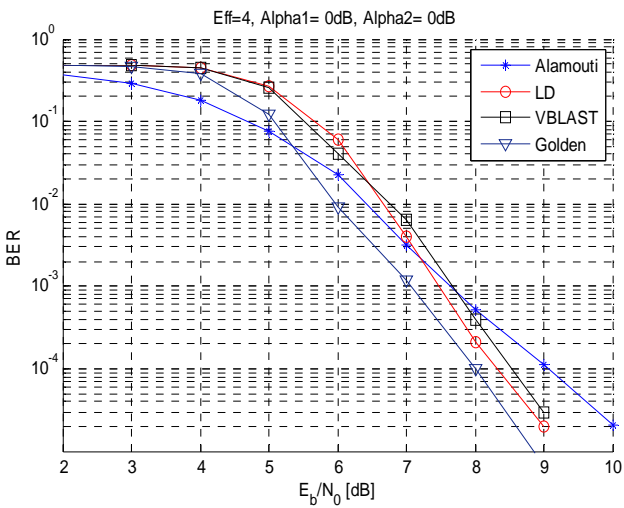

Figure 4- Comparison of different ST coding schemes, Spectral efficiency $\eta$= 4 bit/sec/Hz, 2Rx.

When 3 antennas are used at the receiving side, the conclusions are quite different. Indeed, as shown in Figure 10, Golden code presents the best performance whatever the power attenuation factors on the second and third receiver antennas.

Eventually, we should note from Figure 7 to 10 that the slope of the loss of all NO schemes tends to the same value when the power attenuation factors decrease infinitely. This slope increases with spectral efficiency and decreases with the number of receiving antennas. This means that for 2 receiving antennas a link loss could be completely obtained if the power attenuation factor of the second antenna decreases infinitely. This link loss could be rectified by increasing the number of receiving antennas.

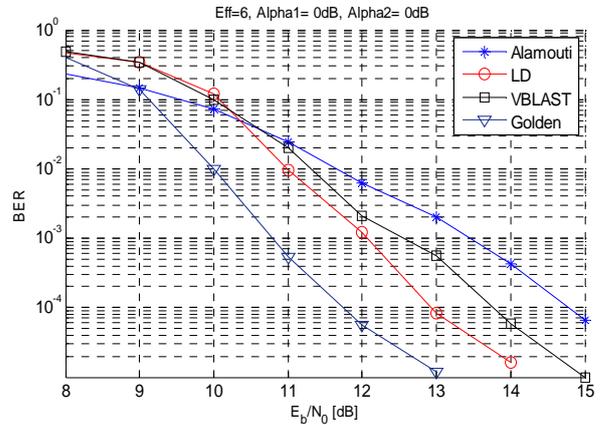

Figure 5- Comparison of different ST coding schemes, Spectral efficiency $\eta$= 6 bit/sec/Hz, 2Rx.

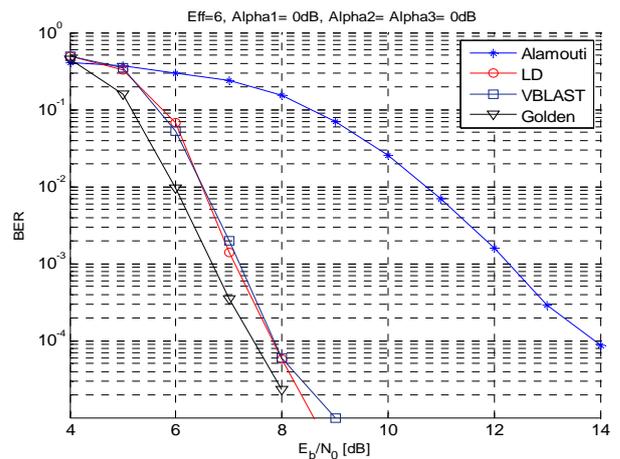

Figure 6- Comparison of different ST coding schemes, Spectral efficiency $\eta$= 6 bit/sec/Hz, 3Rx.

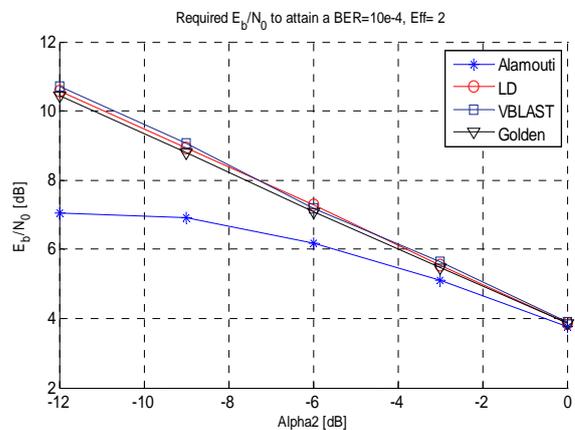

Figure 7- Required $E_b/N_0$ to obtain a BER=$10^{-4}$, Spectral efficiency $\eta$=2 bit/sec/Hz, 2Rx

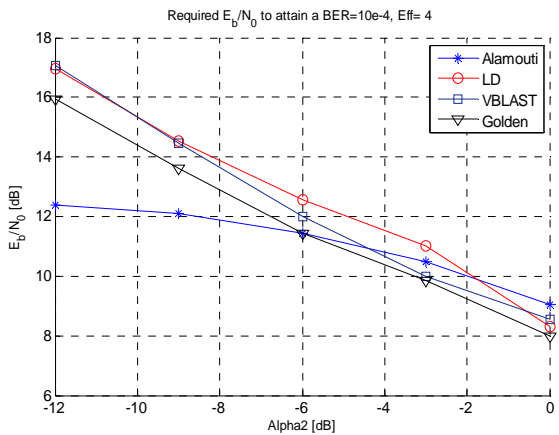

Figure 8- Required $E_b/N_0$ to obtain a BER=$10^{-4}$, Spectral efficiency $\eta$=4 bit/sec/Hz, 2Rx

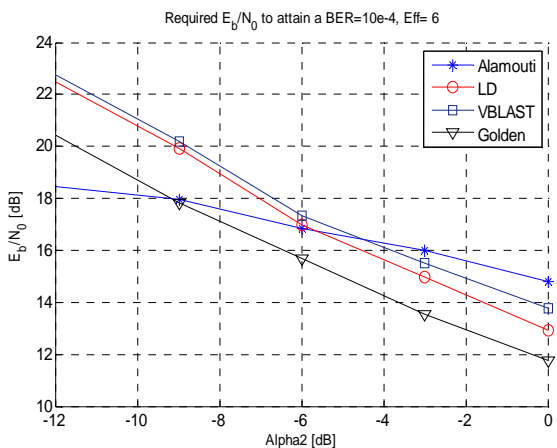

Figure 9- Required $E_b/N_0$ to obtain a BER=$10^{-4}$, Spectral efficiency $\eta$=6 bit/sec/Hz, 2Rx

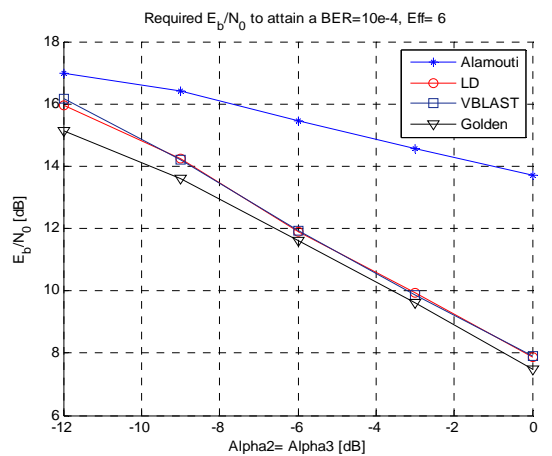

Figure 10- Required $E_b/N_0$ to attain a BER=$10^{-4}$, Spectral efficiency $\eta$=6 bit/sec/Hz, 3Rx

## VI. CONCLUSION

In this paper, we considered the performance of MIMO-OFDM schemes when used with 2 transmitters and 2 or 3 receivers and unbalanced received powers. This study is done using an iterative receiver. We showed by simulations that the convergence of the iterative receiver is obtained after 3 iterations. Moreover, we showed that the superiority of one scheme could not be obtained in all transmission conditions. For 2 receiving antennas, whatever the spectral efficiency is, Alamouti scheme presents the best performance when one antenna is dramatically damaged i.e. when the power received by this antenna decreases infinitely. However, for 3 receivers, Golden code presents the best performance whatever the received powers on different antennas for high spectral efficiencies.

## ACKNOWLEDGMENTS

The authors would like to thank the European CELTIC project "B21C" for its support of this work.